\documentclass[twocolumn,prl,a4paper,aps,showpacs]{revtex4}
\usepackage{graphicx}
\usepackage{dcolumn}
\usepackage{bm}

\begin{document}

\title{Spin-enhanced magnetocaloric effect in molecular nanomagnets}

\author{Marco Evangelisti$^{*}$, Andrea Candini, Alberto Ghirri, and Marco Affronte}
\affiliation{I.N.F.M.$-$S$^{3}$ National Research Center and Dipartimento di Fisica, Universit\`{a} di Modena
e Reggio Emilia, 41100 Modena, Italy}
\author{Euan K. Brechin}
\affiliation{Department of Chemistry, University of Edinburgh, EH9 3JJ Edinburgh, United Kingdom}
\author{Eric J. L. McInnes}
\affiliation{Department of Chemistry, University of Manchester, M13 9PL Manchester, United Kingdom}

\date{\today}

\begin{abstract}
An unusually large magnetocaloric effect for the temperature region below 10~K is found for the
Fe$_{14}$ molecular nanomagnet. This is to large extent caused by its extremely large spin $S$
ground-state combined with an excess of entropy arising from the presence of low-lying excited $S$
states. We also show that the highly symmetric Fe$_{14}$ cluster core, resulting in small cluster
magnetic anisotropy, enables the occurrence of long-range antiferromagnetic order below
$T_{N}=1.87$~K.
\end{abstract}


\maketitle

Nanomagnets are considered good candidates for enhanced magnetocaloric effect (MCE) at low
temperatures, and therefore are of interest for applications as magnetic refrigerants in the low
$T$-range~\cite{parti}. This is mostly because large magnetic moments $S$, resulting therefore in
large magnetic entropies, are attainable in this class of materials. Large $S$, however, is also
often associated with large particle magnetic anisotropy. The larger is the particle magnetic
anisotropy, the higher is the blocking temperature and the lower is the isothermal magnetic
entropy change~\cite{mcemol2}. Ideal materials would rather be nanomagnets with large $S$ and
small anisotropy. Opportunities are provided by molecule-based clusters, which are collections of
identical nanomagnets. Recently, quantum effects were taken into account to explain the MCE of
high-spin molecular clusters, such as Mn$_{12}$ and Fe$_{8}$~\cite{tejada}, whilst chemical
engineering was proposed to enhance MCE in Cr-based molecular rings~\cite{mcemol5}.

In this Letter, we show that the Fe$_{14}$ molecular nanomagnet~\cite{chimica} has a huge MCE in the liquid
helium $T$ region, which is much larger than that of any other known material. We show that this comes out
from a combination of several features, such as the spin ground-state that amounting to $S=25$ is amongst the
highest ever reported, and the highly symmetric cluster core that results in small cluster magnetic
anisotropy. The latter enables the occurrence of long-range magnetic order (LRMO) below $T_{N}=1.87$~K,
probably of antiferromagnetic nature. We also show that low-lying excited $S$ states additionally enhance the
MCE of Fe$_{14}$.

Magnetization measurements down to 2~K and specific heat measurements using the relaxation method
down to $\approx 0.35$~K on powder samples, were carried out in a Quantum Design PPMS set-up for
the $0<H<7$~T magnetic field range. Magnetization and susceptibility measurements below 2~K were
performed using a home-made Hall microprobe appositely installed in the same set-up. In this case
the sample used consisted of a collection of small grains of c.a. $10^{-3}~{\rm mm}^{3}$.

The Fe$_{14}$ molecular cluster, nominally Fe$_{14}$(bta)$_{6}$O$_{6}$(OMe)$_{18}$Cl$_{6}$~\cite{chimica},
has a highly symmetric core in which the Fe$^{3+}$ $s=\frac{5}{2}$ spins are exchange-coupled to each other
by Fe$-$O$-$Fe bridges. Preliminary characterizations~\cite{chimica} and simulations~\cite{mc} have shown
that the Fe$_{14}$ molecule may have a spin ground-state as large as $S=25$ and small cluster magnetic
anisotropy. This is corroborated in Fig.~1 by isothermal magnetization measurements at low-$T$~\cite{oldmag}.
For instance, a fit of $M(H)$ data collected at $T=2$~K provides $S=25$, $g=2.06$ and uniaxial zero-field
splitting as low as $D=0.04$~K. This should be considered a rough estimate of the cluster magnetic
anisotropy, since, as shown below, magnetic data in the liquid helium $T$ region are affected by low-lying
excited $S$ states.

\begin{figure}[b!]
\includegraphics[angle=0,width=5cm]{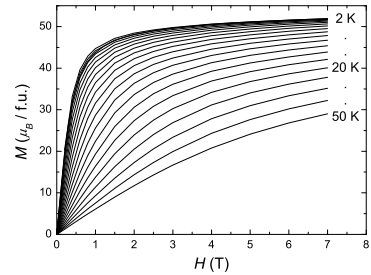}
\caption{Isothermal $M(H)$ curves measured at different temperatures from 2~K to 50~K.}
\end{figure}

Figure~2 shows the magnetic susceptibility $\chi(T)$ and specific heat $C(T)$ data of Fe$_{14}$.
At first look, the main feature is given by the sharp anomaly at $T_{N}=(1.87\pm 0.02)$~K, that
can be seen in both $\chi(T)$ and $C(T)$, and that we attribute to LRMO. The $\chi(T)$ data from
$\sim 10$~K down to 0.35~K, taken with Hall microprobe, are properly scaled with data collected
for $T>2$~K using a calibrated magnetometer, both with applied field $H=0.01$~T. The maximum
$\chi$ at $T_{N}$ corresponds to $\sim 56$~emu/mol (Fig.~2, upper panel), which is smaller than
that expected for paramagnetic $S=25$ spin. This suggests that a full ordered $S=25$ state inside
the cluster is not achieved at $T_{N}$, likely because (i) internal degrees of freedom allow spin
states other than $S=25$ to contribute, and/or (ii) intercluster interactions are similar in
magnitude to the intracluster ones. The observed behavior is compatible with an antiferromagnetic
nature of the ordered phase, as suggested by the sharp decrease of $\chi T(T)$ at low $T$ (inset
of Fig.~2).

\begin{figure}[t!]
\includegraphics[angle=0,width=6cm]{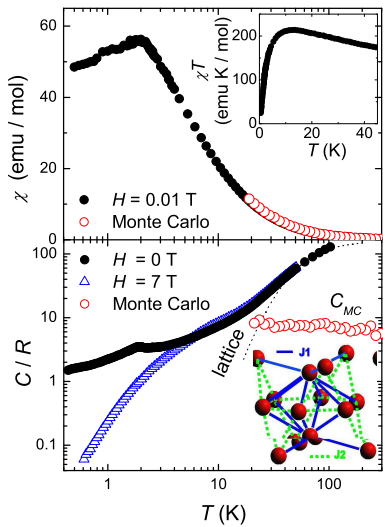}
\caption{(Color online). Top: dc-$\chi(T)$ (and dc-$\chi T(T)$ in the inset) taken for $H=0.01$~T. Bottom:
$C(T)$ for $H=0$ and 7~T. Empty dots are Monte Carlo simulations for $H=0$. Dotted line is the lattice
contribution. The sketch of the metallic core highlights the two chemically distinct exchange interactions
$J_{1}$ (solid line) and $J_{2}$ (dotted line), as indicated.}
\end{figure}

The specific heat data $C/R$, shown in the lower panel of Fig.~2 for $H=0$ and 7~T, corroborate
the interpretation of the $\chi$ data: the $\lambda$-type anomaly in the zero-field $C$ at
$T_{N}$, revealing the onset of LRMO, is quickly removed by the application of an external $H$,
proving its magnetic origin. The occurrence of LRMO implies relatively small cluster magnetic
anisotropy, otherwise superparamagnetic blocking above $T_{N}$ should be observed. However, even a
small anisotropy may become important for a $S$ as large as that of Fe$_{14}$. This is reflected,
for instance, in the relative height of the transition peak at $T_{N}$ that, amounting to $\approx
1.2~R$, is apparently a bit too small for such a large $S$, suggesting that a large portion of the
magnetic entropy is not available for the ordering mechanism. The $C$ data measured above 20~K
show a large increase, that we associate with the lattice contribution~\cite{lattice}.

In what follows, we evaluate the MCE for the Fe$_{14}$ molecular compound from experimental data.
This procedure includes the evaluation of the isothermal magnetic entropy change $\Delta S_{m}$
upon a magnetic field change $\Delta H$, from the measured magnetization and specific heat.
Moreover, we also evaluate the adiabatic temperature change $\Delta T_{ad}$ upon $\Delta H$, from
specific heat data.

In an isothermal process of magnetization, $\Delta S_{m}$ can be derived from Maxwell relations by
integrating over the magnetic field change $\Delta H=H_{f}-H_{i}$, i.e., $\Delta S_{m}(T)_{\Delta
H}=\int_{H_{i}}^{H_{f}}[\partial M(T,H)/\partial T]_{H}~{\rm d}H$. From $M(H)$ data of Fig.~1, the
obtained $\Delta S_{m}(T)$ for $\Delta H=(7-0)$~T~\cite{note} is displayed in the upper panel of
Fig.~3. It can be seen that $-\Delta S_{m}(T)$ reaches a maximum of $4.9~R$ at $T=6$~K.

\begin{figure}[t!]
\includegraphics[angle=0,width=5.3cm]{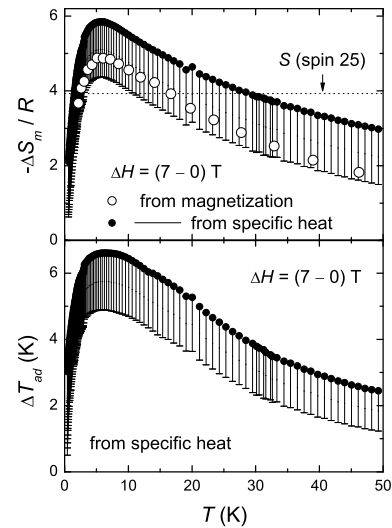}
\caption{Top: $\Delta S_{m}(T)$ as obtained from $C$ (filled dots and bars) and $M$ data (empty dots), both
for $\Delta H=(7-0)$~T. Dotted line is the limiting entropy for $S=25$. Bottom: $\Delta T_{ad}(T)$ as
obtained from $C$ data (filled dots and bars) for $\Delta H=(7-0)$~T.}
\end{figure}

We next turn to the evaluation of MCE from $C$ data of Fig.~2. We firstly determine the total entropies for
$H=0$ and 7~T as functions of $T$, i.e., $S(T)_{H}=\int_{0}^{T}[C(T)_{H}/T]~{\rm d}T$. Experimental entropies
are obtained integrating down to the lowest achieved $T\approx 0.35$~K and, obviously, not from $T=0$~K as
required. To account for the lower-$T$ region, we extrapolate linearly the experimental $C$ below $T_{N}$ for
$T\rightarrow 0$~K, and calculate the associated entropy content. Successively, for $\Delta H=(7-0)$~T, we
calculate $\Delta S_{m}(T)_{\Delta H}=\left[S(T)_{7{\rm T}}-S(T)_{0}\right]_{T}$ and $\Delta
T_{ad}(T)_{\Delta H}=\left[T(S)_{7{\rm T}}-T(S)_{0}\right]_{S}$. Note that the estimation of the lattice
contribution is irrelevant for our calculations, since we deal with differences between total entropies at
different $H$. The results obtained considering the $T\rightarrow 0$~K extrapolation of the experimental $C$,
are displayed in Fig.~3 as filled dots, whereas the added bars are obtained without considering this
extrapolation and can be considered as a lower bound. For $\Delta H=(7-0)$~T and $T=6$~K, we get $-\Delta
S_{m}=(5.0\pm 0.8)~R$, or equivalently $(17.6\pm 2.8)$~J/Kg~K, and $\Delta T_{ad}=(5.8\pm 0.8)$~K. It can be
noticed that, within the bars, the so-obtained $\Delta S_{m}$ fully agrees with the previous estimate
inferred from $M(T,H)$, suggesting that both (independent) procedures can be effectively used to characterize
Fe$_{14}$ with respect to its magnetocaloric properties.

The spin value of Fe$_{14}$ accounts only partially for the large MCE we measured. The experimental $\Delta
S_{m}(T)$ exceeds, indeed, the entropy expected for a $S=25$ spin system, that is $R~{\rm ln}~(2S+1)\simeq
3.9~R$ (Fig.~3). To explain where the observed excess of magnetic entropy change comes from, we model the
magnetic and thermal properties of an isolated Fe$_{14}$ molecule in zero-applied-field by classical Monte
Carlo (MC) simulations using the metropolis algorithm. Following the arguments reported in~\cite{chimica} on
the angles of the Fe$-$O$-$Fe bridges and looking at the bottom inset of Fig.~2, two primary categories of
the Fe$-$O$-$Fe bridges inside the molecule can be identified: those that connect the apical iron ions to the
face cap and equatorial iron ions (whose exchange coupling we indicate as $J_{1}$), and those characterized
by much smaller angles that connect all other iron ions ($J_{2}$). We consider therefore the Hamiltonian
${\mathcal H}=-\sum_{i=1,2}\sum_{(j,k)}J_{i}~\mathbf{s_{j}}\cdot \mathbf{s_{k}}$, for all possible $(j,k)$
pairs of exchange-coupled Fe$^{3+}$ spins. Thus, we calculate $\chi(T)$ to fit the experimental data
(Fig.~2), obtaining estimates for $J_{1}$ and $J_{2}$. To avoid the influence of inter-cluster interactions
and cluster anisotropy, only data for $T>20$~K are taken into account. Assuming $g=2.06$ as deduced from the
saturation of the magnetization, the fit provides $J_{1}/k_{B}\simeq -60.0$~K and $J_{2}/k_{B}\simeq
-25.2$~K, where negative sign indicates that they are both antiferromagnetic. A similar analysis of $\chi(T)$
for Fe$_{14}$ is already reported in Ref.~\cite{mc}. We should mention that, on basis of our MC simulations,
a slight change of the $J_{1}/J_{2}$ ratio has a strong influence on the determination of the cluster spin,
suggesting the competing nature of the exchange interactions inside the molecule. We next use the so-obtained
$J_{i}$ values to calculate the specific heat $C_{MC}$ associated with internal degrees of freedom of the
molecule. We obtain a relatively large contribution $C_{MC}\sim 7$ --- $9~R$ for $20~{\rm K}<T<300$~K
(Fig.~2), that implies the presence of excited states close in energy to the $S=25$ ground-state. Likely,
large $C_{MC}$ values have to be expected in the lower-$T$ region as far as excited states remain populated.
Recalling the uncomplete achievement of the $S=25$ spin state deduced from experimental $\chi(T)$ at $T_{N}$,
and on basis of our MC simulations, we identify the entropy associated with this contribution as that
responsible for the observed enhancement of the MCE of Fe$_{14}$. Additionally, the transition to LRMO is
certainly contributing as well to the MCE parameters below $\sim 2$~K. When LRMO occurs, the magnetization
and magnetic entropy strongly varies in a narrow $T$-range in the vicinity of the transition temperature.
However, on basis of the relatively small height of the ordering peak (Fig.~2), we do not expect this
contribution to be the dominant one. Indeed, no apparent anomaly is seen in the $\Delta S_{m}(T)$ and $\Delta
T_{ad}(T)$ curves at $T_{N}$.

The values of $\Delta S_{m}(T)$ and $\Delta T_{ad}(T)$ obtained in Fe$_{14}$ are exceptionally
large, even more than the ones obtained with intermetallic materials known to be, so far, the best
magnetic refrigerant materials in the $T$ range below 10~K. For instance, the best representative
is the recently studied~\cite{lima1} (Dy$_{x}$Er$_{1-x}$)Al$_{2}$ alloy that, for $x\ge 0.5$
concentrations, presents MCE parameters below 10~K which are at least 30~\% smaller than that of
Fe$_{14}$. Among systems of superparamagnetic particles and molecular magnets, the gap is even
more pronounced. For instance, because of their well-defined spin ground state in this $T$ and $H$
range, it is ease to show that the well known Fe$_{8}$ and Mn$_{12}$-ac molecular nanomagnets,
i.e., cannot exceed values of $-\Delta S_{m}\simeq 12.5$ and 11~J/Kg~K, respectively, thus much
smaller than that of Fe$_{14}$. Moreover in these materials, as in most molecular magnets, an
additional complication (with respect to MCE parameters) is added by the blocking of the cluster
spins in the liquid helium $T$ region, causing the spin-lattice relaxation to slow down
dramatically. Therefore, cluster spins tend to loose thermal contact with the lattice~\cite{marco}
resulting in lower magnetic entropies and, consequently, lower MCE parameters. Ideally, it is
desirable to keep the spin-lattice relaxation at sufficiently high rates down to lowest
temperatures, in order to have a more efficient material in terms of MCE. This route was already
recently tried~\cite{mcemol5} with Cr$_{7}$Cd molecular rings, that can be seen as an ordered
arrangement of well-separated paramagnetic spins, having fast relaxations in the whole
(experimental) $T$ range. In terms of MCE parameters, the only limitation of this material is
given by the low cluster spin value ($S=\frac{3}{2}$) allowing not more than $-\Delta S_{m}\simeq
5.1$~J/Kg~K as experimentally reported to occur for $T<2$~K~\cite{mcemol5}.

Summing up, the above-reported experiments show that the Fe$_{14}$ molecular nanomagnet is unique
in terms of MCE due to the combination of the following characteristics: (i) unusually large spin
ground-state; (ii) small cluster magnetic anisotropy; (iii) excess of entropy resulting from
low-lying excited $S$ states; (iv) long-range magnetic ordering. For these reasons, Fe$_{14}$ has
therefore high potentiality to work as magnetic refrigerant within a temperature range below 10~K.

We thank S. Carretta and N. Magnani for stimulating discussions, and J. F. Fern\'{a}ndez for useful hints on
the MC calculations. This work is partially funded by MIUR under FIRB project no. RBNE01YLKN and by the
EU-Marie Curie network ``QuEMolNa'' and EU-Network of Excellence ``MAGMANet''.


\begin{thebibliography}{99}
\bibitem{oldmag} We should mention here that our magnetization measurements repeatedly provided
results that slightly differ from what reported in Ref.~\cite{chimica}. This discrepancy is beyond the
experimental uncertainty. We have carefully checked the validity of the results here presented and obtained
on fresh samples. We stress that the saturation values of Fig.~1 do agree with $S=25$, as expected from
considerations on the intra-molecular exchange couplings~\cite{chimica}.
\bibitem{lattice} The lattice contribution above 20~K is modelled with $n=55$ optic-like modes having
Einstein temperature $\theta_{E}=68$~K (dotted curve in Fig.~2).
\bibitem{note} For practical reasons, the measurements at the lowest applied field were carried
out for $H_{i}=10^{-3}$~T, which in our calculations was approximated to zero-applied-field.
\end{thebibliography}
\end{document}